\documentclass{llncs}

\newcommand{\couic}[1]{}
\def\lra{\longrightarrow}
\def\lla{\longleftarrow}
\def\vde{\vdash_{\cal R}}

\newbox\tempa
\newbox\tempb
\newdimen\tempc
\def\mud#1{\hfil $\displaystyle{\mathstrut #1}$\hfil}
\def\rig#1{\hfil $\displaystyle{#1}$}
\def\irulehelp#1#2#3{\setbox\tempa=\hbox{$\displaystyle{\mathstrut #2}$}%
		        \setbox\tempb=\vbox{\halign{##\cr
	\mud{#1}\cr
	\noalign{\vskip\the\lineskip}%
	\noalign{\hrule height 0pt}%
	\rig{\vbox to 0pt{\vss\hbox to 0pt{${\; #3}$\hss}\vss}}\cr
	\noalign{\hrule}%
	\noalign{\vskip\the\lineskip}%
	\mud{\copy\tempa}\cr}}%
		      \tempc=\wd\tempb
		      \advance\tempc by \wd\tempa
		      \divide\tempc by 2 }
\def\irule#1#2#3{{\irulehelp{#1}{#2}{#3}%
		     \hbox to \wd\tempa{\hss \box\tempb \hss}}}

\begin{document}

\title{What is a Theory~?}

\author{Gilles Dowek}

\institute{INRIA-Rocquencourt, BP 105, 78153 Le Chesnay Cedex, France.
{\tt Gilles.Dowek@inria.fr} {\tt http://logical.inria.fr/$\tilde{~}$dowek}}

\date{}

\maketitle

\thispagestyle{empty}

\begin{abstract}
{\em Deduction modulo} is a way to express a theory using computation
rules instead of axioms. We present in this paper an extension of
deduction modulo, called {\em Polarized deduction modulo}, where some
rules can only be used at positive occurrences, while others can only
be used at negative ones.  We show that all theories in propositional
calculus can be expressed in this framework and that cuts can always
be eliminated with such theories.
\end{abstract}

Mathematical proofs are almost never built in pure logic, but besides
the deduction rules and the logical axioms that express the meaning of
the connectors and quantifiers, they use something else -~{\em a
theory}~- that expresses the meaning of the other symbols of the
language. Examples of theories are equational theories, arithmetic,
type theory, set theory, ...

The usual definition of a theory, as a set of axioms, is sufficient
when one is interested in the provability relation, but, as
well-known, it is not when one is interested in the structure of
proofs and in the theorem proving process. For instance, we can define
a theory with the axioms $a = b$ and $b = c$ (where $a$, $b$ and $c$
are individual symbols) and prove the proposition $a = c$. However, we
may also define this theory by the computation rules $a \lra b$ and $c
\lra b$ and then a proposition $t = u$ is provable if $t$ and $u$ have
the same normal form using these computation rules. The advantages of this presentation are
numerous.
\begin{itemize}
\item We know that all the symbols occurring in a proof
of $t = u$ must occur in $t$ or in $u$ or one of their reducts. For
instance, the symbol $d$ need not be used in a proof of $a = c$.
We get this way analyticity results.

\item In automated theorem proving, we can use this kind of results
to reduce the search space. In fact, in this case, we just need to
reduce deterministically the terms and check the identity of their
normal forms. We get this way decisions algorithms. 

\item Since the normal form of the proposition $a = d$ is $b = d$ and
$b$ and $d$ are distinct, the proposition $a = d$ is not provable in
this theory. We get this way independence results and, in particular,
consistency results.

\item In an interactive theorem prover, we can reduce the proposition
to be proved, before we display it to the user. This way, the user is
relieved from doing trivial computations.
\end{itemize}

To define a theory with computation rules, not any set of rules is
convenient. For instance, if instead of taking the rules $a \lra b$,
$c \lra b$ we take the rules $b \lra a$, $b \lra c$, we lose the
property that a proposition $t = u$ is provable if $t$ and $u$ have a
common reduct. To be convenient, a rewrite system must be
confluent. Confluence, and sometimes also termination, are necessary
to have analyticity results, completeness of proof search methods,
independence results, ...

When we have rules rewriting propositions directly, for instance
$$x \times y = 0 \lra x = 0 \vee y = 0$$ confluence is not sufficient
anymore to have these results, but cut elimination is also required
\cite{TPM,DowekFrocos}. 
Confluence and cut elimination are related. For instance, with the non
confluent system $b \lra a$, $b \lra c$, we can prove the proposition
$a = c$ introducing a cut on the proposition $b = b$, but, because the
rewrite system is not confluent, this cut cannot be eliminated.
Confluence can thus be seen as a special case of cut elimination when only
terms are rewritten \cite{DowekWollic}, but in the general case, confluence
is not a sufficient condition for cut elimination.

Computation rules are not the only alternative to axioms. Another one
is to add non logical deduction rules to predicate logic either taking
an introduction and elimination rule for the abstraction symbol in
various formulations of set theory
\cite{Prawitz,Crabbe,Hallnas,Bailin,Crabbe91,Ekman} or interpreting
logic programs or definitions as deduction rules
\cite{HallnasSchroederHeister,SchroederHeister92,SchroederHeister93,McDowellMiller}
or in a more general setting \cite{NegriPlato}.  Non logical deduction
rules and computation rules have some similarities, but we believe
that computation rules have some advantages. For instance, non logical
deduction rules may blur the notion of cut in natural deduction and
extra proof reduction rules have to be added (see, for instance,
\cite{Dowek2001}).  Also with some non logical deduction rules, the
contradiction $\bot$ may have a cut free proof and thus consistency
is not always a consequence of cut elimination. In contrast, the notion
of cut, the proof reduction rules and the properties of cut free
proofs remain the usual ones with computation rules.

When a theory is given by a set of axioms, we sometimes want to find
an alternative way to present it with computation rules, in such a way
that cut elimination holds. From cut elimination, we can deduce
analyticity results, consistency and various independence results,
completeness of proof search methods and in some cases decision
algorithms. Many theories have been presented in such a way, including
various equational theories, several presentations of simple type
theory (with combinators or lambda-calculus, with or without the axiom
of infinity, ...), the theory of equality, arithmetic, ...  However, a
systematic way of transforming a set of axioms into a set of rewrite
rules is still to be found. A step in this direction is Knuth-Bendix
method \cite{KB} and its extensions, that permit to transform some
equational theories into rewrite systems with the cut elimination
property (i.e. with the confluence property). Another step in this
direction is the result of S.~Negri and J.~Von Plato \cite{NegriPlato}
that gives a way to transform some sets of axioms, in particular all
quantifier free theories, into a set of non logical deduction rules in
sequent calculus, preserving cut elimination.  In this paper, we
propose a way to transform any consistent quantifier free theory into
a set of computation rules with the cut elimination property.

Our first attempt was to use 
Deduction modulo 
\cite{TPM,DowekWerner} or Asymmetric deduction Modulo \cite{DowekWollic}
as a general framework where computation and deduction can be mixed.
In Deduction modulo, the introduction rule of conjunction 
$$\irule{A~~~B}{A \wedge B}{}$$
is transformed into a rule 
$$\irule{A~~~B}{C}{\mbox{if $C \equiv A \wedge B$}}$$
where $\equiv$ is the congruence generated by the computation
rules, and the other deduction rules are transformed in a similar way.
In Asymmetric deduction modulo, this rule is rephrased 
$$\irule{A~~~B}{C}{\mbox{if $C \lra A \wedge B$}}$$
where the congruence is replaced by the rewriting relation. 

However, although we have no formal proof of it, it seems that the
theory formed with the single axiom $P \Rightarrow Q$ (where $P$ and
$Q$ are proposition symbols) cannot be expressed neither in Deduction
modulo nor in Asymmetric deduction modulo (while the theory $P
\Leftrightarrow Q$ can, as well as more complex theories such as
arithmetic or type theory). Here we shall continue weakening deduction
modulo and introduce {\em Polarized Deduction Modulo} where when
rewriting $C$ to $A \wedge B$ we shall distinguish negative and
positive occurrences of $C$. This way we will be able to transform the
axiom $P \Rightarrow Q$ into the negative rule $P \lra Q$ where $P$
can be rewritten into $Q$ at negative occurrences only, or into the
positive rule $Q \lra P$ where $Q$ can be rewritten into $P$ at
positive occurrences only.

\section{Polarized deduction modulo}

\begin{definition}[Polarized rewrite system]
A {\em rewrite rule} is a pair $P \lra A$ where $P$ is an atomic
proposition and $A$ an arbitrary proposition.
A {\em polarized rewrite system} $\langle {\cal R}_{-}, {\cal R}_{+}
\rangle$ is a pair of sets of rewrite rules. The rules of
${\cal R}_{-}$ are called {\em negative} and those of ${\cal R}_{+}$
are called {\em positive}.
\end{definition}

\begin{definition}[Rewriting]
Given a polarized rewrite system, we define the one step rewriting
relations $\lra^{1}_{-}$ and $\lra^{1}_{+}$ 
\begin{itemize}
\item
if $P \lra A$ is a negative rule then $P \lra^{1}_{-} A$, 

\item
if ($A \lra^{1}_{+} A'$ and $B = B'$)
or 
($A = A'$ and $B \lra^{1}_{-} B'$),\\
then $A \Rightarrow B \lra^{1}_{-} A' \Rightarrow B'$,

\item
if ($A \lra^{1}_{-} A'$ and $B = B'$)
or 
($A = A'$ and $B \lra^{1}_{-} B'$),\\
then $A \wedge B \lra^{1}_{-} A' \wedge B'$ and 
$A \vee B \lra^{1}_{-} A' \vee B'$,
\end{itemize}

\begin{itemize}
\item
if $P \lra A$ is a positive rule then $P \lra^{1}_{+} A$, 

\item
if ($A \lra^{1}_{-} A'$ and $B = B'$)
or 
($A = A'$ and $B \lra^{1}_{+} B'$),\\
then $A \Rightarrow B \lra^{1}_{+} A' \Rightarrow B'$,

\item
if ($A \lra^{1}_{+} A'$ and $B = B'$)
or 
($A = A'$ and $B \lra^{1}_{+} B'$),\\
then $A \wedge B \lra^{1}_{+} A' \wedge B'$ and 
$A \vee B \lra^{1}_{+} A' \vee B'$.
\end{itemize}
Then the rewriting relations $\lra_{-}$ and $\lra_{+}$ are defined as
the transitive closures of the relations
$\lra^{1}_{-}$ and $\lra^{1}_{+}$.
\end{definition}

The deduction rules of {\em Polarized natural deduction modulo} are
those of figure \ref{polarized}. Those of {\em Polarized sequent
calculus modulo} are those of figure \ref{sequent}. 

\begin{figure}
$$
\begin{array}{c}
\irule{}
      {\Gamma \vde B}
      {\mbox{axiom if $A \in \Gamma$ and $A \lra_{-} C \lla_{+} B$}}\\
\irule{\Gamma, A \vde B}
      {\Gamma \vde C}
      {\mbox{$\Rightarrow$-intro if $C \lra_{+} (A \Rightarrow B)$}}\\
\irule{\Gamma \vde C~~~\Gamma \vde A}
      {\Gamma \vde B}
      {\mbox{$\Rightarrow$-elim if $C \lra_{-} (A \Rightarrow B)$}}\\
\irule{\Gamma \vde A~~~\Gamma \vde B}
      {\Gamma \vde C}
      {\mbox{$\wedge$-intro if $C \lra_{+} (A \wedge B)$}}\\
\irule{\Gamma \vde C}
      {\Gamma \vde A}
      {\mbox{$\wedge$-elim if $C \lra_{-} (A \wedge B)$}}\\
\irule{\Gamma \vde C}
      {\Gamma \vde B}
      {\mbox{$\wedge$-elim if $C \lra_{-} (A \wedge B)$}}\\
\irule{\Gamma \vde A}
      {\Gamma \vde C}
      {\mbox{$\vee$-intro if $C \lra_{+} (A \vee B)$}}\\
\irule{\Gamma \vde B}
      {\Gamma \vde C}
      {\mbox{$\vee$-intro if $C \lra_{+} (A \vee B)$}}\\
\irule{\Gamma \vde D~~~\Gamma, A \vde C~~~\Gamma, B \vde C}
      {\Gamma \vde C}
      {\mbox{$\vee$-elim if $D \lra_{-} (A \vee B)$ }}\\
\irule{\Gamma \vde B}
      {\Gamma \vde A}
      {\mbox{$\bot$-elim if $B \lra_{-} \bot$ }}\\
\end{array}
$$
\caption{Polarized natural deduction modulo}
\label{polarized}
\end{figure}

\begin{figure}
$$
\begin{array}{c}
\irule{}
      {A \vde B}
      {\mbox{axiom if $A \lra_{-} C \lla_{+} B$}}\\
\irule{\Gamma, A \vde \Delta ~~~ \Gamma \vde B, \Delta}
      {\Gamma \vde \Delta}
      {\mbox{cut if $A \lla_{-} C \lra_{+} B$}}\\
\irule{\Gamma, B_1, B_2 \vde \Delta}
      {\Gamma, A \vde \Delta}
      {\mbox{contr-left if $A \lra_{-} B_1$, $A \lra_{-} B_2$}}\\
\irule{\Gamma \vde B_1,B_2,\Delta}
      {\Gamma \vde A,\Delta}
      {\mbox{contr-right if $A \lra_{+} B_1$, $A \lra_{+} B_2$}}\\
\irule{\Gamma \vde \Delta}
      {\Gamma, A \vde \Delta}
      {\mbox{weak-left}}\\
\irule{\Gamma \vde\Delta}
      {\Gamma \vde A,\Delta}
      {\mbox{weak-right}}\\
\irule{\Gamma \vde A, \Delta ~~~ \Gamma, B \vde \Delta}
      {\Gamma, C \vde  \Delta}
      {\mbox{$\Rightarrow$-left if $C \lra_{-} (A \Rightarrow B)$}}\\
\irule{\Gamma, A \vde B, \Delta}
      {\Gamma \vde  C, \Delta}
      {\mbox{$\Rightarrow$-right if $C \lra_{+} (A \Rightarrow B)$}}\\
\irule{\Gamma, A, B \vde \Delta}
      {\Gamma, C \vde  \Delta}
      {\mbox{$\wedge$-left if $C \lra_{-} (A \wedge B)$}}\\
\irule{\Gamma \vde A, \Delta~~~\Gamma \vde B, \Delta}
      {\Gamma \vde C, \Delta}
      {\mbox{$\wedge$-right if $C \lra_{+} (A \wedge B)$}}\\
\irule{\Gamma, A \vde \Delta ~~~ \Gamma, B \vde \Delta}
      {\Gamma, C \vde  \Delta}
      {\mbox{$\vee$-left if $C \lra_{-} (A \vee B)$}}\\
\irule{\Gamma \vde A, B, \Delta}
      {\Gamma \vde C, \Delta}
      {\mbox{$\vee$-right if $C \lra_{+} (A \vee B)$}}\\
\irule{}
      {\Gamma, A \vde \Delta}
      {\mbox{$\bot$-left if $A \lra_{-} \bot$}}\\
\end{array}
$$
\caption{Polarized sequent calculus modulo}
\label{sequent}
\end{figure}

As usual, the rules of natural deduction are those of intuitionistic
logic, and the rules for classical logic are obtained by adding the
excluded middle.  The rules of sequent calculus are those of classical
logic and those of intuitionistic logic are obtained by restricting the
right hand part of sequents to have one proposition at most.

For simplicity, we have given only the rules of propositional logic,
but the case of quantifiers is not more complicated.
Notice also that there there is no rule for negation: the proposition
$\neg A$ is an abbreviation for $A \Rightarrow \bot$.

\medskip 

In general, rewriting has two properties. First, it is oriented and
for instance the term $2 + 2$ rewrites to $4$, but the term $4$ does
not rewrite to $2 + 2$.  Then, rewriting preserves provability. For
instance, the proposition $\mbox{\em even}(2 + 2)$ rewrites to
$\mbox{\em even}(4)$ that is equivalent.  Thus we can always transform
the proposition $\mbox{\em even}(2 + 2)$ to $\mbox{\em even}(4)$ and
we never need to backtrack on this operation.

When rewriting is polarized, the first property is kept, but not the
second. For instance, if we have 
the negative rule $P \lra Q$, the sequent $P \vde P$ can be
proved with the axiom rule, but its normal form $Q \vde P$ cannot
(because the positive occurrence of $P$ cannot be rewritten).  Thus,
proof search in polarized deduction modulo may require backtracking on
rewriting.

\medskip 

We shall use a functional notation for proofs as terms.
\begin{definition} (Proof-terms)
Proof-terms are defined inductively as follows.
\begin{eqnarray*}
\pi ::=& ~~&\alpha \\
          &&|~ \lambda \alpha~\pi ~|~ (\pi_{1}~\pi_{2}) \\
         &&|~ \langle \pi_{1}, \pi_{2} \rangle ~|~ \mbox{\it fst}(\pi) ~|~ \mbox{\it snd}(\pi) \\
         &&|~ i(\pi) ~|~ j(\pi) ~|
                  ~ \delta(\pi_{1},\alpha \pi_{2},\beta \pi_{3}) \\
         &&|~ \delta_{\bot}(\pi)
\end{eqnarray*}
\end{definition}

Each proof-term construction corresponds to a natural deduction rule:
terms of the form $\alpha$ express proofs built with the axiom rule,
terms of the form $\lambda \alpha~\pi$ and $(\pi_{1}~\pi_{2})$ express proofs
built respectively with the introduction and elimination rules of the
implication, terms of the form $\langle \pi_{1}, \pi_{2} \rangle$ and
$fst(\pi)$, $snd(\pi)$ express proofs built with the introduction and
elimination rules of the conjunction, terms of the form $i(\pi),
j(\pi)$ and $\delta(\pi_{1},\alpha \pi_{2},\beta \pi_{3})$ express
proofs built with the introduction and elimination rules of the
disjunction, terms of the form $\delta_{\bot}(\pi)$ express proofs built
with the elimination rule of the contradiction.

\section{Proof reduction}

\subsection{Reduction rules}

As in pure logic, a cut in polarized natural deduction modulo is a
sequence formed by an introduction rule followed by an elimination
rule. For instance, the proof
$$\irule{\irule{\irule{\pi_{1}} {\Gamma \vde A} {} ~~~ \irule{\pi_{2}}
{\Gamma \vde B} {} } {\Gamma \vde C} {\mbox{$\wedge$-intro $C \lra_{-}
A \wedge B$}} } {\Gamma \vde A'} {\mbox{$\wedge$-elim $C \lra_{+} A'
\wedge B'$}}$$ is a cut. Eliminating this cut consists in replacing
this proof by the simpler proof $\pi_{1}$. Expressed on proof-terms,
this rule is rephrased
$$\mbox{\it fst}(\langle \pi_{1},\pi_{2} \rangle) \triangleright  \pi_{1}$$
Similar rules can be designed for the other forms of cut, leading to 
the proof rewrite system of figure \ref{reductionrules}.

\begin{figure}
\begin{eqnarray*}
(\lambda \alpha~\pi_{1}~\pi_{2}) &\triangleright& [\pi_{2}/\alpha]\pi_{1}\\
\mbox{\it fst}(\langle \pi_{1},\pi_{2} \rangle) &\triangleright& \pi_{1}\\
\mbox{\it snd}(\langle \pi_{1},\pi_{2} \rangle) &\triangleright& \pi_{2}\\
\delta(i(\pi_{1}),\alpha \pi_{2}, \beta \pi_{3}) 
&\triangleright& [\pi_{1}/\alpha]\pi_{2}\\
\delta(j(\pi_{1}),\alpha \pi_{2}, \beta \pi_{3}) 
&\triangleright& [\pi_{1}/\beta]\pi_{3}
\end{eqnarray*}
\caption{Proof reduction rules}
\label{reductionrules}
\end{figure}

\subsection{Subject reduction}

In the example above, the proof $\pi_{1}$ is a proof of $A$. For the
reduction rule to be correct, we have to make sure that it is also a
proof of $A'$. This is not the case in general, but this is the case
if the relations relations $\lra_{-}$ and $\lra_{+}$ commute, i.e. if
whenever $A \lla_{-} B \lra_{+} C$ then there is a proposition $D$
such that $A \lra_{+} D \lla_{-} C$. Notice that when the relations
$\lra_{-}$ and $\lra_{+}$ are identical, this property is just
confluence.

\begin{proposition}
\label{toto}
If $\lra_{-}$ and $\lra_{+}$ commute, 
$\pi$ is a proof-term of $\Gamma \vde A$ and $A \lra_{-} A'$ 
then $\pi$ is also a proof-term of $\Gamma \vde A'$.
\end{proposition}

\proof{
We prove, by induction on the structure of $\pi$ that, more generally,
if $\pi$ is a proof-term of $\Gamma \vde A$, 
$\Gamma \lra_{+} \Gamma'$ and $A \lra_{-} A'$, 
then $\pi$ is also a proof-term of $\Gamma' \vde A'$.

\begin{itemize}
\item (axiom) If $\pi$ is a variable $\alpha$, we have $B$ in
$\Gamma$ and $B' \lla_{+} B \lra_{-} C \lla_{+} A \lra_{-} A'$ 
thus we have a proposition $D$ such that $B' \lra_{-} D \lla_{+} A'$ 
and $\pi$ is a proof of $\Gamma' \vde A'$.

\item ($\Rightarrow$-intro) If $\pi = \lambda \alpha~\pi_{1}$, then
$\pi_{1}$ is a proof of $\Gamma, B \vde C$ and $A' \lla_{-} A
\lra_{+} B \Rightarrow C$.  Thus there is a proposition $B'
\Rightarrow C'$ such that $A' \lra_{+} B' \Rightarrow C' \lla_{-} B
\Rightarrow C$. Thus $B \lra_{+} B'$ and $C \lra_{-} C'$ and, by
induction hypothesis, $\pi_{1}$ is a proof of $\Gamma', B'\vde
C'$. Thus $\pi$ is a proof of $\Gamma' \vde A'$.

\item ($\Rightarrow$-elim) If $\pi = (\pi_{1}~\pi_{2})$, then
$\pi_{1}$ is a proof of $\Gamma \vde C$ and $C \lra_{-} B
\Rightarrow A$. Thus $C \lra_{-} B \Rightarrow A'$ and, by induction
hypothesis, $\pi_{1}$ is a proof of $\Gamma' \vde B \Rightarrow A'$ and
$\pi_{2}$ is a proof of $\Gamma' \vde B$.  Thus $\pi$ is a proof of
$\Gamma' \vde A'$.

\item ($\wedge$-intro) If $\pi = \langle \pi_{1}, \pi_{2} \rangle$,
then $\pi_{1}$ is a proof of $\Gamma \vde B$ and $\pi_{2}$ is a proof
of $\Gamma \vde C$ and $A' \lla_{-} A \lra_{+} B \wedge C$.  Thus
there is a proposition $B' \wedge C'$ such that $A' \lra_{+} B' \wedge
C' \lla_{-} B \wedge C$.  Thus $B \lra_{-} B'$ and $C \lra_{-} C'$ and,
by induction hypothesis, $\pi_{1}$ is a proof of $\Gamma' \vde B'$ and
$\pi_{2}$ of $\Gamma' \vde C'$.  Thus $\pi$ is a proof of $\Gamma'
\vde A'$.

\item ($\wedge$-elim) If $\pi = \mbox{\it fst}(\pi_{1})$, then
$\pi_{1}$ is a proof of $\Gamma \vde C$ and $C \lra_{-} A \wedge
B$. Thus $C \lra_{-} A' \wedge B$ and, by induction hypothesis,
$\pi_{1}$ is a proof of $\Gamma' \vde A' \wedge B$.  Thus $\pi$ is a
proof of $\Gamma' \vde A'$. The same holds if $\pi = \mbox{\it
snd}(\pi_{1})$.

\item ($\vee$-intro)
If $\pi = i(\pi_{1})$, 
then $\pi_{1}$ is a proof of $\Gamma \vde B$ and 
$A' \lla_{-} A \lra_{+} B \vee C$.  Thus there is a
proposition $B' \vee C'$ 
such that $A' \lra_{+} B' \vee C' \lla_{-} B \vee
C$ and, by induction hypothesis, $\pi_{1}$ is a proof of $\Gamma'
\vde B'$.  Thus $\pi$ is a proof of $\Gamma' \vde A'$.
The same holds if $\pi = j(\pi_{1})$.

\item ($\vee$-elim) If $\pi = \delta(\pi_{1}, \alpha~ \pi_{2}, \beta
\pi_{3})$ then $\pi_{1}$ is a proof of $\Gamma \vde D$, $D \lra_{-}
B \vee C$, $\pi_{2}$ is a proof of $\Gamma, B \vde A$ and $\pi_{3}$
a proof of $\Gamma, C \vde A$.  By induction hypothesis, $\pi_{1}$
is a proof of $\Gamma' \vde B \vee C$, $\pi_{2}$ is a proof of
$\Gamma', B \vde A'$ and $\pi_{3}$ a proof of $\Gamma', C \vde
A'$. Thus $\pi$ is a proof of $\Gamma' \vde A'$.

\item ($\bot$-elim) If $\pi = \delta_{\bot}(\pi_{1})$ then 
$\pi_{1}$ is a proof of $\Gamma \vde B$ and $B \lra_{-} \bot$.  By
induction hypothesis, $\pi_{1}$ is a proof of $\Gamma' \vde \bot$.
Thus $\pi$ is a proof of $\Gamma' \vde A'$.
\end{itemize}
}

\begin{proposition}
\label{totodual}
If $\lra_{-}$ and $\lra_{+}$ commute, 
$\pi$ is a proof-term of $\Gamma \vde A$ and $A \lla_{+} A'$ 
then $\pi$ is also a proof-term of $\Gamma \vde A'$.
\end{proposition}

\proof{
By induction on the structure of $\pi$.
\begin{itemize}
\item (axiom) If $\pi$ is a variable $\alpha$, we have $B$ in
$\Gamma$ and $B \lra_{-} C \lla_{+} A \lla_{+} A'$ 
thus we have $B \lra_{-} C \lla_{+} A'$ and
$\pi$ is a proof of $\Gamma \vde A'$.

\item ($\Rightarrow$-intro) If $\pi = \lambda \alpha~\pi_{1}$, then
$\pi_{1}$ is a proof of $\Gamma, B \vde C$ and 
$A' \lra_{+} A \lra_{+} B \Rightarrow C$. Hence
$A' \lra_{+} B \Rightarrow C$. 
Thus $\pi$ is a proof of $\Gamma \vde A'$.

\item ($\Rightarrow$-elim) If $\pi = (\pi_{1}~\pi_{2})$, then
$\pi_{1}$ is a proof of $\Gamma \vde C$ and $C \lra_{-} B \Rightarrow
A$.  Thus, by proposition \ref{toto}, $\pi_{1}$ is a proof of $\Gamma
\vde B \Rightarrow A$. We have $B \Rightarrow A' \lra_{+} B
\Rightarrow A$. Hence by induction hypothesis $\pi_{1}$ is a proof of
$\Gamma \vde B \Rightarrow A'$.  Thus $\pi$ is a proof of $\Gamma \vde A'$.

\item ($\wedge$-intro) If $\pi = \langle \pi_{1}, \pi_{2} \rangle$,
then $\pi_{1}$ is a proof of $\Gamma \vde B$ and $\pi_{2}$ is a
proof of $\Gamma \vde C$ and $A' \lra_{+} A \lra_{+} B \wedge
C$.  Hence $A' \lra_{+} B \wedge C$.  Thus $\pi$ is a
proof of $\Gamma \vde A'$.

\item ($\wedge$-elim) If $\pi = \mbox{\it fst}(\pi_{1})$, then
$\pi_{1}$ is a proof of $\Gamma \vde C$ and $C \lra_{-} A \wedge
B$. Thus, by proposition \ref{toto}, $\pi_{1}$ is a proof of 
$\Gamma \vde A \wedge B$. We have $A' \wedge B \lra_{+} A \wedge
B$. Hence by induction hypothesis $\pi_{1}$ is a proof of $\Gamma
\vde A' \wedge B$. 
Thus $\pi$ is a proof of $\Gamma \vde A'$. The same holds if $\pi =
\mbox{\it snd}(\pi_{1})$.

\item ($\vee$-intro)
If $\pi = i(\pi_{1})$, 
then $\pi_{1}$ is a proof of $\Gamma \vde B$ and
$A' \lra_{+} A \lra_{+} B \vee C$.  
Hence $A' \lra_{+} B \vee C$.  
Thus $\pi$ is a proof of $\Gamma \vde A'$.
The same holds if $\pi = j(\pi_{1})$.

\item ($\vee$-elim) If $\pi = \delta(\pi_{1}, \alpha~ \pi_{2}, \beta
\pi_{3})$ then $\pi_{1}$ is a proof of $\Gamma \vde D$, $D \lra_{-}
B \vee C$, $\pi_{2}$ is a proof of $\Gamma, B \vde A$ and $\pi_{3}$
a proof of $\Gamma, C \vde A$.  
By induction hypothesis, $\pi_{2}$ is a proof of 
$\Gamma, B \vde A'$ and $\pi_{3}$
of $\Gamma, C \vde A'$. 
Thus $\pi$ is a proof of $\Gamma \vde A'$. 

\item ($\bot$-elim) If $\pi = \delta_{\bot}(\pi_{1})$ then 
$\pi_{1}$ is a proof of $\Gamma \vde B$, $B \lra_{-} \bot$.  By
proposition \ref{toto}, $\pi_{1}$ is a proof of $\Gamma \vde \bot$.
Thus $\pi$ is a proof of $\Gamma \vde A'$.
\end{itemize}
}

\begin{proposition} (Subject reduction)
If $\lra_{-}$ and $\lra_{+}$ commute, $\pi$ is a proof of $\Gamma \vde
A$ and $\pi \triangleright \pi'$ then $\pi'$ is a proof of $\Gamma \vde A$.
\end{proposition}

\proof{
By induction over the length of the reduction. For the one step case,
we consider the different cases according to the form of the redex.
\begin{itemize}
\item 
If $\pi = (\lambda \alpha \pi_{1}~\pi_{2})$, then $\pi_{1}$ is
a proof of $\Gamma, B' \vde A'$.
The term $\lambda \alpha \pi_{1}$ is a proof of $\Gamma \vde C$
with $C \lra_{+} B' \Rightarrow A'$.
The term $\pi_{2}$ is a proof of $\Gamma \vde B$ and 
$(\lambda \alpha \pi_{1}~\pi_{2})$
is a proof of $A$ with 
with $C \lra_{-} B \Rightarrow A$. 
By commutation, we have $B' \Rightarrow A' \lra_{-} B'' \Rightarrow A''
\lla_{+} B \Rightarrow A$.  Thus $B' \lra_{+} B'' \lla_{-} B$ and $A'
\lra_{-} A'' \lla_{+} A$.
By propositions \ref{toto} and \ref{totodual}, $\pi_{1}$ is a proof of
$\Gamma, B' \vde A$ and $\pi_{2}$ is a proof of $\Gamma \vde B'$.
Thus $\pi'$ is a proof of $\Gamma \vde A'$.

\item
If $\pi = \mbox{\it fst}(\langle \pi_{1}, \pi_{2} \rangle)$, then $\pi_{1}$ is
a proof of $\Gamma \vde A'$, and $\pi_{2}$ a proof of $\Gamma \vde 
B'$. The term $\langle \pi_{1}, \pi_{2} \rangle$ is a proof of $\Gamma
\vde C$
with $C \lra_{+} A' \wedge B'$ and 
$\mbox{\it fst}(\langle \pi_{1}, \pi_{2} \rangle)$
is a proof of $\Gamma \vde A$ with  
$C \lra_{-} A \wedge B$. 
By commutation, there is a proposition $A'' \wedge B''$ such that
$A' \wedge B' \lra^{-} A'' \wedge B''\lla^{+} A \wedge B$.
Hence 
$A' \lra^{-} A'' \lla^{+} A$
and by propositions \ref{toto} and \ref{totodual}, $\pi'$ is a proof
of $A$. 
The same holds if $\pi = \mbox{\it snd}(\langle \pi_{1}, \pi_{2} \rangle)$.

\item
If $\pi = \delta(i(\pi_{1}), \pi_{2}, \pi_{3})$, then $\pi_{1}$ is
a proof of $\Gamma \vde B$, 
the term $i(\pi_{1})$ is a proof of $\Gamma \vde D$
with $D \lra_{+} B \vee C$,
the term
$\pi_{2}$ is a proof of $\Gamma, B' \vde A$ and 
$\pi_{3}$ a proof of $\Gamma, C' \vde A$ 
with $D \lra_{-} B' \vee C'$.
By commutation, there is a proposition $B'' \vee C''$ such that
$B \vee C \lra^{-} B'' \vee C'' \lla^{+} B' \vee C'$.
Hence $B \lra^{-} B'' \lla^{+} B'$
and by propositions \ref{toto} and \ref{totodual}, 
$\pi_{1}$ is a proof of $\Gamma \vde B'$
Thus $\pi'$ is a proof of $A$.
The same holds if $\pi = \delta(j(\pi_{1}), \pi_{2}, \pi_{3})$.
\end{itemize}
}

\subsection{Termination}

This section is an adaptation to polarized deduction modulo of the cut
elimination proof, {\em \`a la} Tait and Girard, of
\cite{DowekWerner}.

\begin{definition} (Neutral proof)

A proof is said to be {\em neutral} if its last rule is an axiom or an
elimination, but not an introduction.
\end{definition}

\begin{definition} (Reducibility candidate)

A set $R$ of proofs is a {\em reducibility
candidate} if 
\begin{itemize}
\item if $\pi \in R$, then $\pi$ is strongly normalizable,
\item if $\pi \in R$ and $\pi \triangleright \pi'$ then $\pi' \in R$,
\item if $\pi$ is neutral and 
if for every $\pi'$ such that $\pi \triangleright^{1} \pi'$, $\pi' \in R$ then 
$\pi \in R$. 
\end{itemize}

Let ${\cal C}$ be the set of all reducibility candidates.
\end{definition}

\begin{definition} (Pre-model)
Consider a language ${\cal L}$, a {\em pre-model} for ${\cal L}$ is a
function associating a reducibility candidate $\hat{P}$ to each atomic
proposition $P$.
\end{definition}

\begin{definition}
Let $A$ be a proposition.
We define the set $|A|$ of proofs by induction over the
structure of $A$.

\begin{itemize}
\item
If $P$ is atomic then $|P| = \hat{P}$.

\item
A proof $\pi$ is element of $|A \Rightarrow B|$ if it
is strongly normalizable and when $\pi$ reduces to a proof of the form
$\lambda \alpha \pi_{1}$ then for every $\pi'$ in $|A|$, 
$[\pi' / \alpha]\pi_{1}$ is an element of $|B|$.

\item
A proof $\pi$ is an element of $|A \wedge B|$ if it is strongly 
normalizable and when $\pi$ reduces to a proof of the form
$\langle \pi_{1},\pi_{2} \rangle$ then $\pi_{1}$ and $\pi_{2}$ are elements of 
$|A|$ and $|B|$.

\item
A proof $\pi$ is an element of $|A \vee B|$ if it is strongly 
normalizable and when $\pi$ reduces to a proof of the form
$i(\pi_{1})$ (resp. $j(\pi_{2})$) then $\pi_{1}$ (resp. $\pi_{2}$) is
an element of $|A|$ (resp. $|B|$). 

\item
A proof $\pi$ is an element of $|\bot|$ if it is strongly
normalizable.
\end{itemize}
\end{definition}

\begin{proposition} 
For every proposition $A$, $|A|$ is a reducibility candidate.
\end{proposition}

\proof{See \cite{DowekWerner}.}

\medskip

In deduction modulo, a pre-model is a pre-model of a rewrite system
${\cal R}$, if for each rule $P \lra A$, we have $|P| = |A|$. In the
polarized case, we take the following weaker condition.

\begin{definition}
\label{defpremodelof}
A pre-model is a pre-model of a polarized rewrite system ${\cal R}$ if
\begin{itemize}
\item for each negative rule $P \lra A$, we have $|P| \subseteq |A|$,
\item for each positive rule $P \lra A$, we have $|A| \subseteq |P|$.
\end{itemize}
\end{definition}

\begin{proposition}
Let ${\cal R}$ be a polarized rewrite system, in a pre-model of 
${\cal R}$ we have 
\begin{itemize}
\item if $A \lra_{-} B$ then $|A| \subseteq |B|$,
\item if $A \lra_{+} B$ then $|B| \subseteq |A|$.
\end{itemize}
\end{proposition}

\proof{By induction over the structure of $A$.}

\begin{theorem}
Let ${\cal R}$ be a polarized rewrite system such that $\lra_{-}$ and 
$\lra_{+}$ commute and that has a pre-model. Let $A$ be a proposition,
$\pi$ be a proof of $A$ modulo ${\cal 
R}$ and $\sigma$ a substitution mapping proof variables of
propositions $B$ to elements of $|B|$. Then $\sigma \pi$ is an element
of $|A|$.
\end{theorem}

\proof{By induction over the structure of $\pi$. For sake of brevity,
we detail only the cases of the axiom rule and the rules of
implication.

\begin{itemize}
\item (axiom)
If $\pi$ is a variable $\alpha$, then $\sigma \pi = \sigma \alpha$.
If $\alpha$ is bound by $\sigma$ then $\sigma \pi$ is an element of
$|C|$ and $|C| \subseteq |B| \subseteq |A|$. Hence $\sigma \pi \in |A|$. 

If $\alpha$ is not bound by $\sigma$ then $\sigma \pi = \alpha$ and
thus it is in $|A|$.

\item ($\Rightarrow$-intro)
The proof $\pi$ has the form $\lambda \alpha \rho$ where
$\alpha$ is a proof variable of some proposition $B$ and $\rho$ a
proof
of some proposition $C$.
We have $\sigma \pi = \lambda \alpha \sigma \rho$, 
consider a reduction sequence issued from this proof. This sequence
can only reduce the proof $\sigma \rho$. 
By induction hypothesis, the proof $\sigma \rho$ is an
element of $|C|$, thus the reduction sequence is finite. 

Furthermore, every reduct of $\sigma\pi$ is of the form 
$\lambda \alpha \rho'$ where $\rho'$ is a reduct of $\sigma \rho$. 
Let then $\tau$ be any proof of $|B|$, the proof $[\tau /
\alpha]\rho'$
can be obtained by reduction from $([\tau / \alpha] \circ
\sigma) \rho$. By induction hypothesis, the proof 
$([\tau / \alpha] \circ \sigma) \rho$ is an element of 
$|C|$. Hence, as $|C|$ is a reducibility candidate,
the proof $[\tau / \alpha]\rho'$ is an element of $|C|$.

Hence, the proof $\sigma \pi$ is an element of $|B \Rightarrow C|$. 
As $A \lra_{+} B \Rightarrow C$, we have $|B \Rightarrow C| \subseteq
|A|$, hence $\sigma \pi \in |A|$.

\item ($\Rightarrow$-elim)
The proof $\pi$ has the form $(\rho_{1}~\rho_{2})$ and $\rho_{1}$ is a
proof of some proposition $C$ such that $C \lra_{-} B \Rightarrow A$
and $\rho_{2}$ a proof of
the proposition $B$. We have $\sigma \pi = (\sigma \rho_{1}~\sigma
\rho_{2})$. By induction hypothesis $\sigma \rho_{1}$ and $\sigma
\rho_{2}$ are in the sets $|C|$ and $|B|$. 
As $C \lra_{-} B
\Rightarrow C$ we have $|C| \subseteq |B \Rightarrow A|$ and thus
$\sigma \rho_{1} \in |B \Rightarrow A|$.
Hence these
proofs are strongly normalizable. Let $n$ be the maximum length of a
reduction sequence issued from $\sigma \rho_{1}$ and $n'$ the maximum
length of a reduction sequence issued from $\sigma \rho_{2}$. We
prove by induction on $n + n'$ that $(\sigma \rho_{1}~\sigma
\rho_{2})$ is in the set $|A|$. As this proof is neutral we only need
to prove that every of its one step reducts is in $|A|$. If the
reduction takes place in $\sigma \rho_{1}$ or in $\sigma \rho_{2}$
then we apply the induction hypothesis. Otherwise $\sigma \rho_{1}$
has the form $\lambda \alpha~\rho'$ and the reduct is $[\sigma
\rho_{2}/\alpha]\rho'$. By the definition of $|B \Rightarrow A|$ this
proof is in $|A|$.

Hence, the proof $\sigma \pi$ is an element of $|A|$. 

\end{itemize}
}

\begin{corollary}
Every proof of $A$ is in $|A|$ and hence strongly normalizable
\end{corollary}

Using the same technique as in \cite{DowekWerner} we can 
extend this cut elimination result to intuitionistic polarized sequent calculus
modulo, provided we extend the proof reduction rules with the
ultra-reduction rules
$$\delta(\pi_{1},\alpha\pi_{2},\beta\pi_{3}) \triangleright \pi_{2}$$
$$\delta(\pi_{1},\alpha\pi_{2},\beta\pi_{3}) \triangleright \pi_{3}$$
We can also extend the result to classical sequent calculus, 
defining a classical pre-model of a rule $P \lra A$ as a pre-model of
the rule $P \lra A''$ where $A''$ is the light double negation of $A$
defined by
\begin{itemize}
\item $A'' = A$ if $A$ is atomic,
\item $(A \Rightarrow B)'' = A' \Rightarrow B'$,
\item $(A \wedge B)'' = A' \wedge B'$,
\item $(A \vee B)'' = A' \vee B'$,
\item $\bot'' = \bot$,
\end{itemize}
with
\begin{itemize}
\item $A' = \neg \neg A$ if $A$ is atomic,
\item $(A \Rightarrow B)' = \neg \neg (A' \Rightarrow B')$,
\item $(A \wedge B)' = \neg \neg (A' \wedge B')$,
\item $(A \vee B)' = \neg \neg (A' \vee B')$,
\item $\bot' = \neg \neg \bot$.
\end{itemize}

\section{The equivalence Lemma}

We are now ready to relate theories defined by axioms and by polarized
rewrite rules.

\begin{proposition}[Equivalence]
Let ${\cal R}$ be a polarized rewrite system
such that $\lra_{-}$ and $\lra_{+}$ commute.
Let ${\cal T}$ be the
set of axioms formed with, for each negative rule $P \lra
A$ of ${\cal R}$ the axiom $P \Rightarrow A$ and for each positive
rule $P \lra A$ of ${\cal R}$ the axiom $A \Rightarrow P$.
Then
$$\Gamma \vde A$$
if and only if
$${\cal T}, \Gamma \vdash A$$
\end{proposition}

\proof{Notice, first, that for every proposition $A \Rightarrow B$ of
${\cal T}$, the sequent $A \vde B$ is provable with the axiom rule,
using either the rule $A \lra_{-} B$ or the rule $B \lra_{+} A$ and
thus the sequent $\vde A \Rightarrow B$ is provable.
Using propositions \ref{toto} and \ref{totodual} for every proposition
$D$ such that $A \Rightarrow B \lra_{-} C \lla_{+} D$, the sequent
$\vde D$ is provable. 
Thus, by induction over the structure of a proof of ${\cal T}, \Gamma
\vdash A$, we build a proof of $\Gamma \vde A$ replacing by a proof all
invocations to the axioms of ${\cal T}$.

Conversely, we first prove, by induction over the structure of $A$
that if $A \lra_{-} B$ then ${\cal T} \vdash A \Rightarrow B$
and that if $A \lra_{+} B$ then ${\cal T} \vdash B \Rightarrow A$.
Thus, by induction over the structure of a proof of
$\Gamma \vde A$, we build a proof of ${\cal T}, \Gamma \vdash A$.
As an example, we give the case of the
$\wedge$-intro rule. The proof has the form
$$\irule{\irule{\pi_{1}}
               {\Gamma \vde B}
               {}
         ~~~
         \irule{\pi_{2}}
               {\Gamma \vde C}
               {}
        }
        {\Gamma \vde A}
        {\mbox{$\wedge$-intro~~~where $A \lra_{+} B \wedge C$}}$$
By the induction hypothesis we have proofs $\pi_{1}'$ and $\pi_{2}'$ of 
${\cal T}, \Gamma \vdash B$ and ${\cal T}, \Gamma \vdash C$. We first build
the proof: 
$$\irule{\irule{\pi_{1}'}
               {{\cal T}, \Gamma \vdash B}
               {}
         ~~~
         \irule{\pi_{2}'}
               {{\cal T}, \Gamma \vdash C}
               {}}
        {{\cal T}, \Gamma \vdash B \wedge C}
        {\mbox{$\wedge$-intro}}$$
Then, we have $A \lra_{+} (B \wedge C)$, thus ${\cal T} \vdash (B
\wedge C) \Rightarrow A$ and we can build a proof of the proposition $A$.}

\section{Expressing axioms as rewrite rules}

Now, we consider a theory given by a set of quantifier free axioms and
we want to present it with a polarized rewrite system such that
$\lra_{-}$ and $\lra_{+}$ commute and such that polarized deduction
modulo this rewrite system has the cut elimination property.

In polarized deduction modulo, there is no cut free proof of $\bot$.
Thus cut elimination implies consistency and consistency is a
necessary condition for a set of axioms to be transformed into a
polarized rewrite system.  As we shall see, this condition is
sufficient. We shall prove that any consistent theory can be presented
with a polarized rewrite system such that the left hand sides of the
negative rules and positive rules are disjoint (i.e. no atomic
proposition can be rewritten both by a negative and a positive
rule). We shall also see that with such a rewrite system the relations
$\lra_{-}$ and $\lra_{+}$ always commute and that cut elimination
always holds.

\begin{proposition}
Consider a rewrite system such that the left hand sides of the
negative rules and positive rules are disjoint, then the relation 
$\lra_{-}$ and $\lra_{+}$ commute and cut elimination holds.
\end{proposition}

\proof{Commutation is a simple consequence of the absence of critical
pairs. To prove cut elimination, we build a pre-model. Left hand sides
of negative rules by the smallest reducibility candidate (the
intersection of all reducibility candidates) 
and all the other atomic propositions by the largest (the set of all
strongly normalizable proof-terms). The conditions of definition 
\ref{defpremodelof} hold obviously, thus cut elimination holds for 
intuitionistic natural deduction. This results extends to 
intuitionistic sequent calculus and classical sequent calculus.}

\begin{theorem}
\label{thetheo}
If a quantifier free theory is consistent, then it can be presented as
a rewrite system such that the relation $\lra_{-}$ and $\lra_{+}$
commute and cut elimination holds.
\end{theorem}

\proof{Let $\Gamma$ be a consistent set of quantifier free axioms.
We prove that $\Gamma$ can be presented as
a rewrite system such that the left hand sides of the negative rules
and positive rules are disjoint.

Let $\nu$ be a model of $\Gamma$.
Following \cite{NegriPlato}, we consider 
the conjunctive-disjunctive (clausal form) $\Gamma'$ of $\Gamma$. 

We pick a clause of $\Gamma'$. In this clause there is either a literal
of the form $P$ such that $\nu(P) = 1$ or a literal of the form $\neg
P$ such that $\nu(P) = 0$.

In the first case, we pick all the clauses of $\Gamma'$ where $P$ occurs
positively 
$$P \vee A_{1}, ..., P \vee A_{n}$$
we replace these clauses by the proposition
$$(\neg A_{1} \vee ... \vee \neg A_{n}) \Rightarrow P$$
and then by the positive rule 
$$P \lra_{+} \neg A_{1} \vee ... \vee \neg A_{n}$$

In the second, we pick all the clauses of $\Gamma'$ where $P$ occurs
negatively 
$$\neg P \vee A_{1}, ..., \neg P \vee A_{n}$$
we replace these clauses by the proposition
$$P \Rightarrow (A_{1} \wedge ... \wedge A_{n})$$
and then by the negative rule 
$$P \lra_{-} A_{1} \wedge ... \wedge A_{n}$$

We repeat this process with all clauses of $\Gamma'$. We obtain this
way a polarized rewrite system ${\cal R}$.  All the rules have a left
hand side whose interpretation in $\nu$ is $1$ if the rule is positive
and $0$ if the rule is negative. Hence the left hand sides of the
negative rules and positive rules are disjoint.}

\begin{example}
In deduction modulo, it is well-known that consistency does not imply
cut elimination. For instance, the theory defined by the rewriting rule 
(Crabb\'e's rule)
$$A \lra (B \wedge \neg A)$$
is consistent but does not have the cut elimination property
\cite{Crabbe,DowekWerner}.

However this rule is not the only way to present this theory in
deduction modulo and the algorithm above can be used to find another
presentation. 
The proposition $A \Leftrightarrow (B \wedge \neg A)$ has a model 
$\nu(A) = \nu(B) = 0$ and from its clausal form 
$$(\neg A) \wedge (\neg A \vee B) \wedge (A \vee \neg B)$$
we get the rules 
$$A \lra_{-} \bot$$
$$B \lra_{-} A$$
\end{example}

Notice that in this case, the theory can be also presented with the
simpler, non polarized rewrite system $A \lra \bot$, $B \lra A$.
It remains to be investigated which theories can be expressed with a
non polarized system and which ones cannot.

\section*{Conclusion}

With any proof search method (such as resolution or tableaux), a lot
of work is duplicated when we search for a proof of a proposition $A$
and then of a proposition $B$ in the same axiomatic theory $\Gamma$.
This suggests that before searching for proofs in some theory, we
should ``prepare'' the theory and express it in such a way that this
duplication is avoided. This preparation of a theory can be compared
to the compilation of a program: a program is compiled once and not
each time it is executed.

An extreme case is when the theory $\Gamma$ is contradictory, proof
search in the prepared theory should then be trivial.  When the theory
is consistent, the search for a proof in the prepared theory should
restrict to analytic proofs, i.e. the search for a proof of a
proposition $A$ should involve only the sub-formulas of $A$ (including
instances and reducts) and thus the search of a proof of the
contradiction $\bot$ should fail immediately.

Transforming an axiomatic theory into a rewrite system such that
deduction modulo this rewrite system has the cut elimination property
is an example of such a preparation. Knuth-Bendix method permits to
do this for some equational theories. We have proposed here a similar
preparation method for quantifier free theories. Of course, the
general case still needs to be investigated.

Since preparation seems to involve a consistency check, and
consistency is undecidable in general, such a preparation method
can only be partial.

At last, it is well-known that when a theory has the cut elimination
property, then it is consistent, but that the converse does not hold:
there are consistent theories that do not have the cut elimination
property, for instance set theory or the theory $A \lra B \wedge \neg
A$ \cite{Crabbe,DowekWerner}. However, we have seen that various
presentations of the same theory may or may not have the cut
elimination property. Thus, if we take the definition that an
axiomatic theory has the cut elimination property if {\em one} of its
presentation in deduction modulo has the cut elimination property,
then the theory $A \lra B \wedge \neg A$ has the cut elimination
property and the problem is open for set theory (while \cite{Bailin}
seems to suggest that the result might be positive).  Then, it is not
so obvious that there are consistent theories that do not have the cut
elimination property, and in particular we have seen that for
quantifier free theories, consistency and cut elimination coincide.
The generality of this result also remains to be investigated.

\section*{Acknowledgements}

I want to thank Th\'ere\`se Hardin, H\'el\`ene Kirchner, Claude
Kirchner and Benjamin Werner for many discussions on deduction modulo, 
cut elimination and Knuth-Bendix method.


\begin{thebibliography}{99.}

\bibitem{Bailin}
S.C.~Bailin. 
A normalization theorem for set theory. 
{\em The Journal of Symbolic Logic}, 53, 3, 1988, pp.~673-695.

\bibitem{Crabbe}
M.~Crabb\'e.
Non-normalisation de la th\'eorie de {Z}ermelo.
Manuscript, 1974.

\bibitem{Crabbe91}
M.~Crabb\'e. 
Stratification and cut-elimination.
{\em The Journal of Symbolic Logic}, 56, 1991, pp.~213-226.

\bibitem{DowekFrocos}
G. Dowek. Axioms vs. rewrite rules: from completeness to cut
elimination. H.~Kirchner and Ch.~Ringeissen (Eds.), {\em Frontiers of
Combining Systems}, Lecture Notes in Artificial Intelligence 1794,
Springer-Verlag, 2000, pp.~62-72.

\bibitem{Dowek2001}
G. Dowek. About folding-unfolding cuts and cuts modulo. {\em Journal of
Logic and Computation} 11, 3, 2001, pp.~419-429.  

\bibitem{DowekWollic}
 G. Dowek. Confluence as a cut elimination property. 
{\em Workshop on Logic, Language, Information and Computation}, 2001.

\bibitem{TPM}
G.~Dowek, Th.~Hardin, and C.~Kirchner. Theorem proving modulo. 
{\em Journal of Automated Reasoning} (to appear).
{\em Rapport de Recherche INRIA} 3400, 1998.

\bibitem{DowekWerner}
G.~Dowek and B.~Werner.
Proof normalization modulo.
{\em Types for proofs and programs}, T.~Altenkirch,
W.~Naraschewski, and B.~Rues (Eds.), {\em Lecture Notes in Computer Science}
1657, Springer-Verlag, 1999, pp.~62-77. 
{\em Rapport de Recherche} 3542, INRIA, 1998.

\bibitem{Ekman}
J. Ekman. {\em Normal proofs in set theory}. Doctoral
thesis, Chalmers University of Technology and University of
G\"{o}teborg, 1994.

\bibitem{Hallnas}
L. Halln\"{a}s. {\em On normalization of proofs in set theory}. 
Doctoral thesis, University of Stockholm, 1983.

\bibitem{HallnasSchroederHeister}
L.~Halln\"{a}s and P.~Schroeder-Heister. A proof-theoretic approach to
logic programming. I. Clauses as rules. {\em Journal of Logic and
Computation} 1, 2, 1990, pp.~261-283.  II. Programs as
definitions. {\em Journal of Logic and Computation} 1, 5, 1991,
pp.~635-660. 

\bibitem{KB}
D.E.~Knuth and P.B.~Bendix.
Simple word problems in universal algebras.
J.~Leech (Ed.), {\em Computational Problems in Abstract
Algebra}, Pergamon Press, 1970, pp.~263-297. 


\bibitem{McDowellMiller}
R.~McDowell and D.~Miller. 
Cut-Elimination for a logic with definitions and induction. 
{\em Theoretical Computer Science} 232, 2000, pp.~91-119.

\bibitem{NegriPlato}
S.~Negri and J.~Von Plato. Cut elimination in the presence of axioms. 
{\em The Bulletin of Symbolic Logic}, 4, 4, 1998, pp.~418-435.

\bibitem{Prawitz}
D.~Prawitz.
{\em Natural deduction, a proof-theoretical study}.
Almqvist \& Wiksell, 1965.

\bibitem{SchroederHeister92}
P.~Schroeder-Heister. Cut elimination in logics with definitional
reflection. 
D.~Pearce and H.~Wansing (Eds.), {\em Nonclassical Logics and Information 
Processing}, Lecture Notes in Computer Science 619, 
Springer-Verlag, 1992, pp.~146-171.

\bibitem{SchroederHeister93}
P.~Schroeder-Heister. Rules of definitional reflection. {\em Logic in
Computer Science}, 1993, pp.~222-232.

\end{thebibliography}
\end{document}